\documentclass[epj,final]{svjour}
\usepackage{graphicx, bm, mcite, paralist}

 \setdefaultleftmargin{0em}{4mm}{}{}{}{}
\begin{document}

\title{Far-infrared optical excitations in multiferroic TbMnO$_{3}$}
%\subtitle{Spin-phonon coupling in multiferroic TbMnO$_3$}

\author{M.~Schmidt\inst{1}, Ch.~Kant\inst{1},
T.~Rudolf\inst{1}, F.~Mayr\inst{1}, A.A.~Mukhin\inst{2}, A.M.~Balbashov\inst{3}, J. Deisenhofer\inst{1}, \and
A.~Loidl\inst{1} }

\institute{Experimental Physics V, Center for Electronic Correlations and Magnetism, University of Augsburg,
D-86135~Augsburg, Germany \and General Physics Institute of the Russian Academy of Sciences, 119991 Moscow, Russia
\and Moscow Power Engineering Institute, 105835 Moscow, Russia}

\date{\today}

\abstract{We provide a detailed study of the reflectivity of multiferroic TbMnO$_3$ for wave numbers from
40~cm$^{-1}$ to 1000~cm$^{-1}$ and temperatures 5~K $<$ \emph{T} $<$ 300~K. Excitations are studied for
polarization directions \textbf{E}${}\parallel{}$\textbf{a}, the polarization where electromagnons are observed,
and for \textbf{E}${}\parallel{}$\textbf{c}, the direction of the spontaneous polarization in this material. The
temperature dependencies of eigenfrequencies, damping constants and polar strengths of all modes are studied and
analyzed. For \textbf{E}${}\parallel{}$\textbf{a} and below the spiral ordering temperature of about 27~K we
observe a transfer of optical weight from phonon excitations to electromagnons, which mainly involves
low-frequency phonons. For \textbf{E}${}\parallel{}$\textbf{c} an unusual increase of the total polar strength and
hence of the dielectric constant is observed indicating significant transfer of dynamic charge probably within
manganese-oxygen bonds on decreasing temperatures.}

%\abstract{We provide a detailed study of the reflectivity of multiferroic TbMnO$_3$ for wave numbers 40~cm$^{-1}$
%to 1000~cm$^{-1}$ and temperatures 5~K $<$ T $<$ 300~K. Excitations are studied for polarization directions
%\textbf{E}${}\parallel{}$\textbf{a}, the polarization where electromagnons are observed, and for
%\textbf{E}${}\parallel{}$\textbf{c}, the direction of the spontaneous polarization in this material. The
%temperature dependencies of eigenfrequencies, damping constants and polar strengths of all modes are studied and
%analyzed. For \textbf{E}${}\parallel{}$\textbf{a} and below the spiral ordering temperature of about 27~K we
%observe a clear transfer of optical weight from phonon excitations to electromagnons, which mainly involves
%low-frequency phonons. For \textbf{E}${}\parallel{}$\textbf{c} an unusual increase of the total polar strength and
%hence of the dielectric constant is observed indicating significant transfer of dynamic charge probably within
%manganese-oxygen bonds on decreasing temperatures.}

%\doi{}


\titlerunning{Excitations in multiferroic TbMnO$_3$}
\authorrunning{M. Schmidt {\it et al.}}
\maketitle

\section{Introduction}

The discovery of spin-driven ferroelectricity in some rare earth manganites \cite{Kimura2003}  has revived the
field of multiferroics enormously \mcite{Fiebig2005,*Cheong2007,*Ramesh2007}. In these systems the onset of
ferroelectricity is directly coupled to the static non-collinear spin order \mcite{Katsura2005, Mostovoy2006,
Sergienko2006}. Concerning the dynamics of multiferroics, the idea of hybridization between spin waves and optical
phonons in magnetoelectrics has been put forth long ago by Smolenskii and Chupis \cite{Chupis1982}, and indeed,
ample experimental evidence for the occurrence of electromagnons in the rare earth manganites has been provided
recently \cite{Pimenov2006, Sushkov2007, Aguilar2007, Pimenov2008, Sushkov2008, Pimenov2008a}. The origin and the
underlying mechanisms of these magnetoelectric excitations have been debated vividly
\mcite{Takahashi2008,*Kida2008}, however, recent studies \cite{Aguilar2009, Pimenov2009} support the scenario that
magnons in spin-driven ferroelectrics can gain optical weight via the coupling to phonons and, hence, can be
observed directly in far-infrared (FIR) and THz spectroscopy. A strong correlation of electromagnons and phonons
has been already reported for the case of GdMnO$_3$ as function of temperature and magnetic field
\cite{Pimenov2006a}.

TbMnO$_3$ is the far most studied multiferroic manganite and crystallizes in an orthorhombically distorted
perovskite structure (space group \emph{Pbnm}). It reveals magnetic phase transitions into an antiferromagnetic
(AFM) collinear spin structure at $T_\mathrm{N} = 42$~K and subsequently in a structure with helical spin order
and spontaneous ferroelectric (FE) polarization at $T_{\mathrm{FE}}= 27$~K \cite{Kimura2003}. In the low
temperature structure the spins reveal helicoidal order with a propagation vector along the crystallographic
\textbf{b} direction and the spin spiral plane lying within the bc plane \cite{Kenzelmann2005}. In this case the
FE polarization is expected to be along the \textbf{c} direction \cite{Katsura2007, Mostovoy2006} in agreement
with experiment \cite{Kimura2003}. In TbMnO$_3$ electromagnons have been observed at wave numbers close to
20~cm$^{-1}$ \cite{Pimenov2006, Aguilar2009, Pimenov2009} and 60~cm$^{-1}$ \cite{Aguilar2009}, which exactly
matches spin wave dispersion energies as determined in neutron scattering experiments
\mcite{Senff2007,*Senff2008}.

For the low-energy electromagnon the optical eigenfrequency corresponds to the new magnetic Bragg point at
(0,0.28,1) and the magnetoelectric excitation at 60~cm$^{-1}$ matches the magnon energy at the magnetic-zone
boundary \mcite{Aguilar2009, Senff2007,*Senff2008}. Recently it has been argued that this range of electromagnons
strongly overlaps with a continuum band of two-magnon absorption \mcite{Takahashi2008,*Kida2008}. It seems,
however, that the two-magnon scattering is too weak to explain the experimentally observed electromagnon
intensities \cite{Aguilar2009}. For the low frequency mode, it has been clearly documented in optical experiments
using THz spectroscopy that the observed electromagnon eigenfrequencies correspond to spin waves, as AFM resonance
absorptions have been detected using different excitation geometries in the same frequency range
\cite{Pimenov2009}. In TbMnO$_3$ it also became clear that the electromagnons only appear for
\textbf{E}${}\parallel{}$\textbf{a} and are strictly tied to the lattice. Furthermore it has been shown that
Heisenberg coupling is very important \cite {Aguilar2009} and that the dynamical magnetoelectric effect as
proposed by Katsura, Nagaosa and Balatsky can play no dominant role in TbMnO$_3$ \cite{Katsura2007}.

The aim of this work is to investigate the temperature dependence of all infrared active phonon modes in TbMnO$_3$
in order to reveal the coupling of phonons and spin waves and to analyze the transfer of optical spectral weight
at the onset of the spiral-spin order.

\section{Experimental details}

The investigated single crystals have previously been used for studies in the THz frequency range
\cite{Pimenov2006, Pimenov2009}. The samples were characterized by x-ray, magnetic, thermodynamic and dielectric
measurements and exhibited good agreement with published results \cite{Kimura2003, Kimura2005}. The optical
experiments were carried out using Bruker Fourier transform spectrometers IFS 113v and IFS 66v/S equipped with a
He-flow cryostat for wave numbers from 40~cm$^{-1}$ to 32,000~cm$^{-1}$ and for temperatures from 5~K to room
temperature. The measurements were performed for light polarized parallel to the crystallographic $\mathbf{c}$
axis ($\mathbf{E}\parallel \mathbf{c}$), the direction of the FE polarization, and for $\mathbf{E}\parallel
\mathbf{a}$, the polarization where electromagnons are observable.

\begin{figure}[b]
\includegraphics[width=0.9\columnwidth]{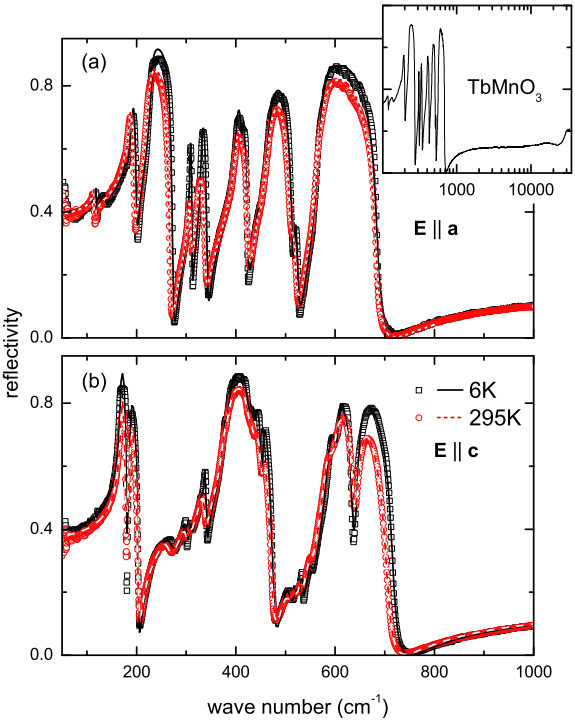}
\caption{(Color online) Reflectivity of TbMnO$_3$ at 295~K (open red circles) and 6~K (open black squares) for
\textbf{E}${}\parallel{}$\textbf{a} (upper frame) and \textbf{E}${}\parallel{}$\textbf{c} (lower frame). The inset
shows the experimentally observed reflectivity up to 32,000~cm$^{-1}$ for \textbf{E}${}\parallel{}$\textbf{a} at
room temperature. The corresponding fit curves are shown as red dashed lines for 295~K and black solid lines at
6~K.} \label{fig1}
\end{figure}

The reflectivity spectra obtained at 6~K and 295~K for $\mathbf{E}\parallel \mathbf{a}$ and $\mathbf{E}\parallel
\mathbf{c}$ are shown in Fig.~\ref{fig1}(a) and Fig.~\ref{fig1}(b), respectively. The reflectivity has been fitted
with RefFIT \cite{kuzmenko05}, using a sum of Lorentzian oscillators with eigenfrequency, damping and dielectric
strength as relevant fitting parameters \cite{Note}.
%\footnote{Note that the excitation M$_1$ was not included in the fit
%routine, because it is very close to the experimental low-frequency limit and a band-fourier-filter was used to
%smoothen the interference fringes superimposed on the spectra due to multiple reflections in the sample.}

The eigenfrequencies correspond to the transverse optical phonon modes and the longitudinal mode frequencies have
to be calculated via the Lyddane-Sachs Teller relation. From the dielectric strength the effective ionic plasma
frequencies can be determined, which allows some conclusions concerning the effective dynamic charges involved in
specific modes \cite{Rudolf2007}.

For clarity in some cases we also show the frequency dependence of the dielectric loss. The complex dielectric
constant has been derived from the reflectivity spectra by means of Kramers-Kronig transformation with a constant
extrapolation towards low frequencies and a smooth $\omega^n$ high-frequency extrapolation.

The inset of Fig.~\ref{fig1} shows the reflectivity of TbMnO$_3$ with \textbf{E}${}\parallel{}$\textbf{a} at room
temperature as measured up to 32,000~cm$^{-1}$ ($\approx$ 4~eV). The hump close to 16,000~cm$^{-1}$ probably
corresponds to dipole forbidden onsite transitions within the \emph{d} multiplet of the manganese ions
\mcite{Yamauchi2009,*Tobe2001,*Kovaleva2004}. The charge transfer gap appears close to 24,000~cm$^{-1}$,
corresponding to 3~eV. The analysis of these electronic transitions will not be subject of this work. In the
present work the high-energy reflectivity is merely used to determine the electronic dielectric constant
$\epsilon_\infty$ with high accuracy.\\

\section{Experimental results and discussion}

In the orthorhombic perovskite structure of TbMnO$_3$ (space group $Pbnm$) one expects a total of 25 IR active
modes, namely 9 B$_{3u}$ for $\mathbf{E}\parallel \mathbf{a}$, 9 B$_{2u}$ modes for $\mathbf{E}\parallel
\mathbf{b}$ and 7 B$_{1u}$ modes for $\mathbf{E}\parallel \mathbf{c}$. However, when fitting the reflectivity
spectra we used 14 modes to describe the spectra for $\mathbf{E}\parallel \mathbf{a}$ and 19 modes for
$\mathbf{E}\parallel \mathbf{c}$. The corresponding fits describe the data nicely and are shown in
Fig.~\ref{fig1}.

\begin{figure}[t]
\centering
\includegraphics[width=0.9\columnwidth]{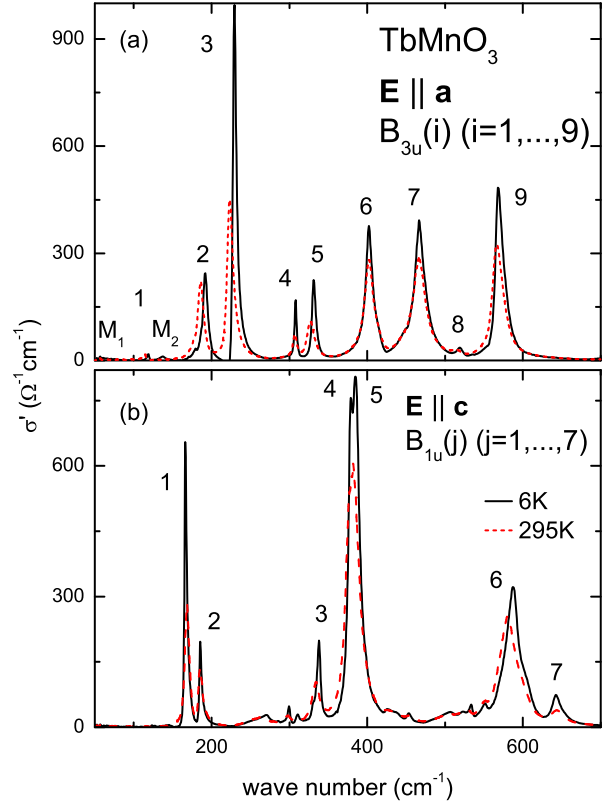}
\caption{(Color online) Optical conductivity of TbMnO$_3$ at 295~K (red dashed line) and 6~K (black solid line)
for \textbf{E}${}\parallel{}$\textbf{a} (a: upper frame) and \textbf{E}${}\parallel{}$\textbf{c} (b: lower
frame).} \label{fig2}
\end{figure}

To identify the main IR active modes we plot the optical conductivity in Fig.~\ref{fig2}(a)
(\textbf{E}${}\parallel{}$\textbf{a}) and Fig.~\ref{fig2}(b) (\textbf{E}${}\parallel{}$\textbf{c}) for
temperatures at 6~K (solid lines) and 295~K (dashed lines). In these figures we assigned the modes with the
largest optical spectral weight to the expected modes B$_{3u}$(i) (i $=1,...,9$) and B$_{1u}$(j) (j $=1,...,7$)
for the $Pbnm$ symmetry. The additional weak modes which are visible are thought to originate from phonon modes
allowed for light polarized along one of the other crystallographic axis due to polarization leakage or a slight
misorientation of the sample.

%Specifically for the polarization direction E${}\parallel{}$\textbf{c} (Fig.~\ref{fig2}(b)) a number of small loss
%peaks appears in the spectrum which hardly can be identified. If these peaks correspond to the leakage of phonon
%excitations symmetry allowed in other directions or to two phonon processes is unclear at present.

%The upper frame (a) of Fig.~\ref{fig2} shows the loss for modes excited along the \textbf{a}-direction. The
%phonon modes are numbered from 1 to 9, which should read P1 to P9 and should indicate all infrared allowed
%B$_{3u}$ modes. In this assignment we ignored small intensities which appear e.g. close to 180~cm$^{-1}$ in the
%low-frequency wing of phonon mode 2 or close to 450~cm$^{-1}$.

The weak excitations labeled M$_1$ and M$_2$ which appear only for \textbf{E}${}\parallel{}$\textbf{a} are related
with the appearance of multiferroicity in this system and, hence, we will now first focus on the FIR spectra for
this polarization direction.

\begin{figure}
\centering
\includegraphics[width=0.98\columnwidth]{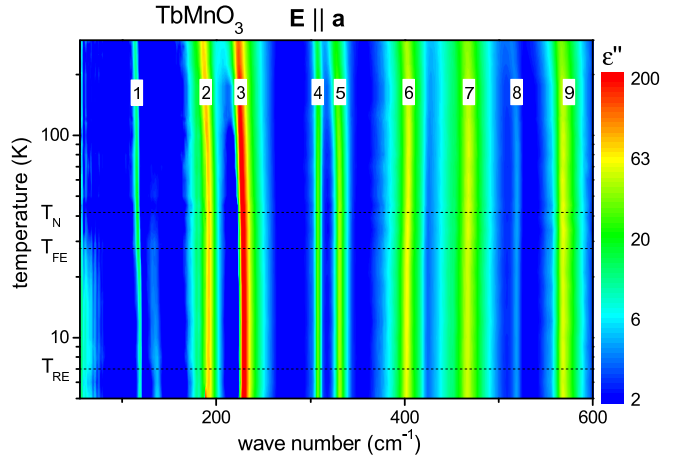}
\caption{(Color online) Color coded contour plot of the dielectric loss as function of temperature vs. wave number
in TbMnO$_3$ for \textbf{E}${}\parallel{}$\textbf{a}. Strong loss contributions are indicated in red. The color
code extends from values of the dielectric loss $\epsilon''=200$ (red) to $\epsilon''=2$ (blue).} \label{Cplot}
\end{figure}

\begin{figure}[t]
\includegraphics[width=\columnwidth]{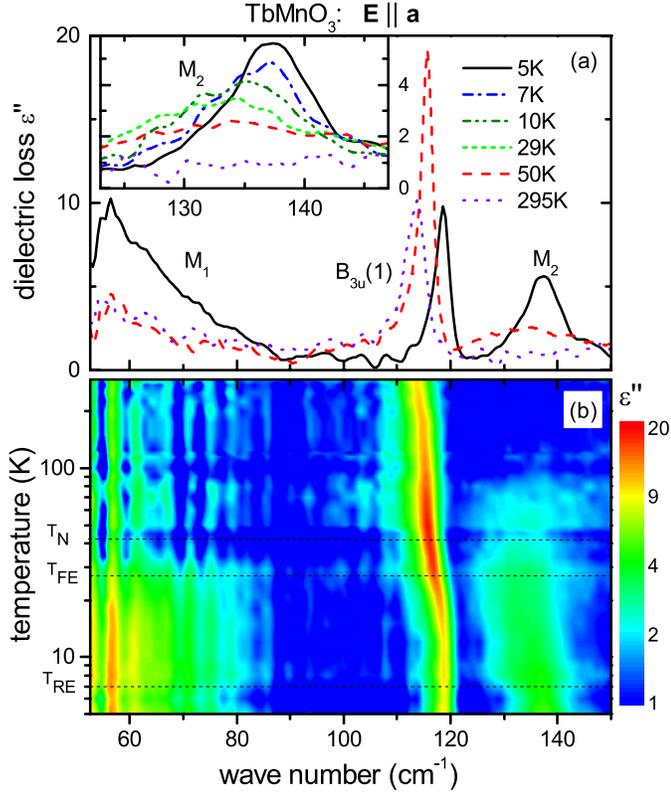}
\caption{\textbf{(a)}(Color online) Upper panel: Dielectric loss of TbMnO$_3$ for
\textbf{E}${}\parallel{}$\textbf{a} between 50 and 150~cm$^{-1}$ at a series of temperatures as indicated in the
figure. The modes M$_1$ and M$_2$ are magnetic excitations. Mode B$_{3u}$(1) is the phonon mode which couples most
strongly to the magnetic excitations. The inset shows the evolution of M$_2$ with temperature.
 \textbf{(b)} Color coded contour plot of the dielectric
loss as function of temperature and wave number for the same range of frequencies.  The color code follows a
logarithmic scale and spans the range between 1 (blue) and 20 (red).} \label{fig3}
\end{figure}

The overall temperature dependence of the FIR spectrum for \textbf{E}${}\parallel{}$\textbf{a} is illustrated  in
a contour plot of the dielectric loss $\epsilon''$ as a function of temperature and wave number (see
Fig.~\ref{Cplot}). In this representation all phonon modes are visible, but no abrupt anomalies at the phase
transition temperatures can be detected. At the temperature $T_\mathrm{RE} = 7$~K magnetic ordering of the Tb ions
takes place. The effects of the magnetic and ferroelectric phase transitions on the FIR spectra become evident
when zooming into the low-frequency region between 50 and 150~cm$^{-1}$, which is shown in Fig.~\ref{fig3}.

The excitations M$_1$ and M$_2$ appearing close to 60 and 140~cm$^{-1}$, respectively, emerge and gain optical
weight only below about 40~K (see Fig.~\ref{fig3}(b)), although some diffusive bands seem to persist to higher
temperatures. For M$_2$ this behavior is documented in the inset of Fig.~\ref{fig3}(a). Therefore, these modes are
associated with the onset of magnetic ordering at $T_\mathrm{N}=42$~K. In contrast, all phonon modes and here
especially B$_{3u}$(1) show finite optical weight up to room temperature. Recently, the excitation M$_1$ has been
identified as an electromagnon, i.e.~an electric-dipole active excitation of a single zone boundary magnon
\cite{Aguilar2009}. Note that several electric-dipole active electromagnons and magnetic-dipole active
antiferromagnetic resonances have been observed by Pimenov et al. in the frequency region below 25~cm$^{-1}$ in
TbMnO$_3$ using THz spectroscopy \cite{Pimenov2006, Pimenov2009}.

To date the nature and origin of M$_2$ remain unsettled. It has been argued that this excitation is due to a
two-magnon process (M$_1$+M$_1$)  with an energy corresponding to the upper cut-off of the magnon density of
states \cite{Takahashi2008}. Recently, this scenario has been discarded in favor of an isotropic
Heisenberg-coupling mechanism between non-collinear spins which allows to explain that electromagnons occur only
for \textbf{E}${}\parallel{}$\textbf{a} \cite{Aguilar2009}. A further possibility is an excitation of the
crystal-field split ground state of the 4$f$ electrons of the Tb ions. However, the relation to the experimentally
observed temperature dependence and to the magnetic ordering of the Mn ions is not in favor of such an
interpretation. When looking at the temperature evolution of the lowest-lying phonon B$_{3u}$(1) close to
120~cm$^{-1}$, one recognizes that upon entering into the ferroelectric phase below $T_{\mathrm{FE}}=27$~K the
phonon shifts and loses intensity (see Fig.~\ref{fig3}(a) and (b)) in agreement with reference
\cite{Takahashi2008}. It is this mode which reportedly transfers optical weight to the electromagnons in GdMnO$_3$
\cite{Pimenov2008a}. Turning again to Fig.~\ref{fig3} it seems that
 both B$_{3u}$(1) and M$_2$ shift to slightly higher frequencies below $T_{\mathrm{FE}}$, but the difference
$\Delta \omega=\omega_{\mathrm{M}_2}-\omega_{\mathrm{B}_{3u}(1)}\cong$ 20~cm$^{-1}$ remains almost constant and in
agreement with the direct excitation energy of the low-energy electromagnon and antiferromagnetic resonances at
20~cm$^{-1}$ \cite{Pimenov2009,Pimenov2006}. Thus, we suggest to consider the possibility that M$_2$ could arise
from a one-magnon + one-phonon process (1M1P). Multi-magnon plus phonon processes have been described in detail in
antiferromagnetic Cu and Ni oxides \cite{Lorenzana1995}, however, only excitations of even numbers of magnons will
yield considerable spectral weight and conserve the total spin. If 1M1P excitations should be realized in
TbMnO$_3$, spin-orbit coupling and antisymmetric Dzyaloshinsky-Moriya-like spin coupling ought to contribute to
the effective spin-dependent electric-dipole moment.

%To shed some more light on this problem, Fig.~\ref{fig2} shows the dielectric loss for both directions and for the
%same range of energies. The upper frame (a) of Fig.~\ref{fig2} shows the loss for modes excited along the
%\textbf{a}-direction. The phonon modes are numbered from 1 to 9, which should read P1 to P9 and should indicate
%all infrared allowed B$_{3u}$ modes.

%Fig.~\ref{fig2}~b shows the phonon modes with \textbf{E}${}\parallel{}$\textbf{c}. According to the symmetry allowed 7
%B$_{1u}$ modes we assigned 7 modes, which are the dominating modes in the loss spectrum. But we have to have in
%mind that we used 19 modes to fit the reflectivity. And indeed a number of small loss peaks appear just below 330~cm$^{-1}$ and between 410 and 570~cm$^{-1}$. We speculate that these modes correspond to leakage modes or two
%phonon excitations, although the temperature dependence is not in favor of anharmonic effects. In this
%polarization it is even less obvious if mode 7, which is rather weak, is a symmetry allowed mode.

%It is worthwhile to mention that we observe no new phonon modes at the AFM phase transition, indicating that no
%symmetry breaking occurs in the lattice. At the onset of antiferromagnetic order the high-temperature symmetry is
%broken and usually new phonon modes appear in the magnetically ordered state. In addition, in frustrated magnets
%it is well established that some phonon modes undergo significant splitting even in the absence of a clear
%symmetry lowering as observed in high-resolution diffraction experiments. Experimentally a splitting of phonon
%modes has been observed in a number of strongly frustrated spinel compounds \mcite{Sushkov2005,
%Rudolf2007b,*Hemberger2006,*Hemberger2007,*Rudolf2007a} and has been explained in terms of a spin-driven
%Jahn-Teller effect \mcite{Yamashita2000,*Tchernyshyov2002}. \\

%The main aim of this work is to provide a detailed analysis of the temperature dependence of characteristic phonon
%parameters. This will be done in the following.

\begin{figure*}
\centering
\includegraphics[width=16cm]{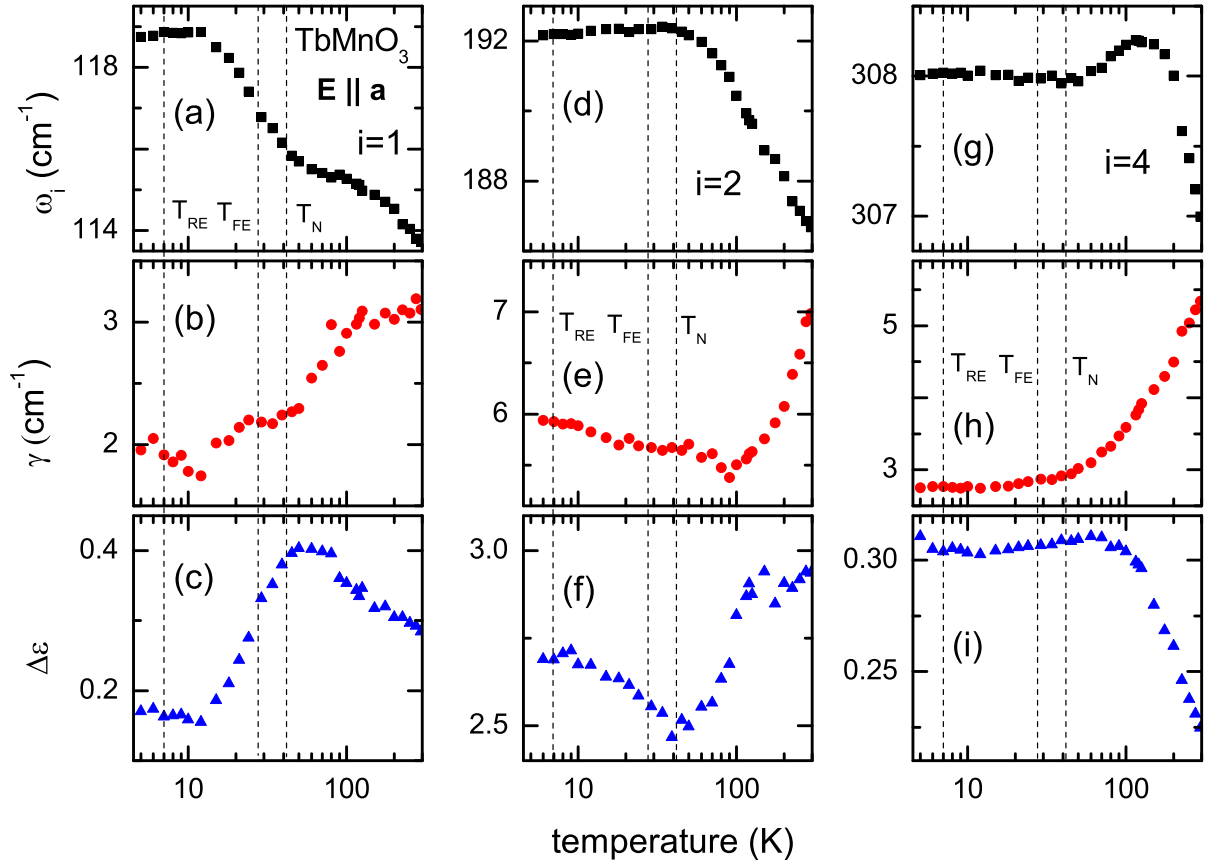}
\caption{(Color online) Temperature dependencies of eigenfrequency (a, d, g), damping constant (b, e, h) and
dielectric strength (c, f, i) for modes B$_{3u}$(i) (i $=1,2,4$) in TbMnO$_3$ for
\textbf{E}${}\parallel{}$\textbf{a} on semilogarithmic scale. The magnetic ordering temperatures are indicated by
vertical dashed lines.} \label{fig5}
\end{figure*}

For a more quantitative analysis of the phonon modes we analyzed all spectra utilizing a 3-parameter fit as
outlined above. Table~\ref{tab1} shows eigenfrequencies $\omega_0$ for both polarization directions at room
temperature and also indicates the experimentally observed dipole strengths $\Delta\epsilon$ and calculated
effective ionic plasma frequencies $\Omega$, which directly provide a measure of the effective dynamic charge
involved in a given mode. The effective ionic plasma frequency of all modes, which characterizes the ionicity of
the bonds amounts to approximately 1100~cm$^{-1}$. It has to be compared with the effective ionic plasma frequency
$\Omega = 2100$~cm$^{-1}$, which results from a model assuming ideal ionic bonds (see e.g.
Ref.~\cite{Rudolf2007}). Let us now turn to the polarization \textbf{E}${}\parallel{}$\textbf{a}, the polarization
in which the electromagnons show up. Figure~\ref{fig5} documents the temperature dependence of eigenfrequency
$\omega$, damping constant $\gamma$ and
dielectric strength $\Delta\epsilon$ of phonon modes B$_{3u}$(i) with i $=1,2,4$.\\

\begin{table}[b]
\centering \caption{Phonon eigenfrequencies $\omega_0$, ionic plasma frequencies $\Omega$ and dipolar strengths
$\Delta\epsilon$ of all 7 B$_{1u}$ and 9 B$_{3u}$ modes in TbMnO$_3$ at room temperature.} \label{tab1}
\begin{tabular}{rrrr}
\multicolumn{1}{l}{\textbf{B$_{1u}$}}     &   \multicolumn{3}{l}{\textbf{7 Phonons}}\\
\\
\multicolumn{1}{c}{Mode}      &   \multicolumn{1}{l}{$\omega_0$ [cm$^{-1}$]} & \multicolumn{1}{l}{$\Omega$
[cm$^{-1}$]} & \multicolumn{1}{l}{$\Delta\epsilon$}\\

\cline{1-4}\noalign{\smallskip}
1    &   169.5    &   301.7   &  3.19  \\
2    &   186.7    &   233.5   &  1.54  \\
3    &   335.4    &   176.5   &  0.35  \\
4    &   377.7    &   395.1   &  1.09  \\
5    &   384.9    &   705.4   &  3.44  \\
6    &   580.4    &   589.0   &  0.99  \\
7    &   644.4    & 168.4     &  0.07  \\

\cline{1-4}\noalign{\smallskip}

\multicolumn{2}{l}{$\Sigma$}     &   1098   &   10.67\\

\\
\\

\multicolumn{1}{l}{\textbf{B$_{3u}$}}     &   \multicolumn{3}{l}{\textbf{9 Phonons}}\\
\\
\multicolumn{1}{c}{Mode}      &   \multicolumn{1}{l}{$\omega_0$ [cm$^{-1}$]} & \multicolumn{1}{l}{$\Omega$
[cm$^{-1}$]} &   \multicolumn{1}{l}{$\Delta\epsilon$}\\

\cline{1-4}\noalign{\smallskip}
1   &   113.7   &   60.7   &   0.28\\
2   &   186.7   &  319.9   &   2.94\\
3   &   225.0   &  494.9   &   4.84\\
4   &   307.0   &  145.6   &   0.22\\
5   &   327.4   &  257.1   &   0.62\\
6   &   401.8   &  467.4   &   1.35\\
7   &   466.5   &  569.0   &   1.49\\
8   &   513.9   &   76.8   &   0.02\\
9   &   567.7   &  550.1   &   0.94\\
\cline{1-4}\noalign{\smallskip}
\multicolumn{2}{l}{$\Sigma$}  &  1135   &   12.7\\
\end{tabular}
\end{table}

Looking now at mode B$_{3u}$(1) which seems to be most strongly coupled to spin excitations we find the following
temperature dependence: Coming from room temperature its eigenfrequency (Fig.~\ref{fig5}(a)) reveals the usual
hardening due to anharmonicity and starts to saturate below 100~K. We tentatively ascribe this temperature scale
to the Curie-Weiss temperature of the manganese subsystem, because in spinel systems spin-phonon coupling effects
became evident below the corresponding Curie-Weiss temperatures \mcite{Sushkov2005,
Rudolf2007b,*Hemberger2006,*Hemberger2007,*Rudolf2007a,*Rudolf2009}. Moreover the temperature scale of 100~K
agrees well with the Curie-Weiss temperature of LaMnO$_3$ \cite{Paraskevopoulos2000}. But on further cooling and
passing the magnetic phase transitions the eigenfrequency suddenly increases further. Experimentally, it seems
that the increase starts already below the first antiferromagnetic phase transition where collinear spin order is
established and further increases when passing the phase transition into a ferroelectric ground state and helical
spin order. In total, this increase due to the onset of AFM order amounts to 3\% indicating strong spin-phonon
coupling. The damping (Fig.~\ref{fig5}(b)) continuously decreases without significant traces of the magnetic and
ferroelectric phase transitions. Finally, the most striking effects show up in the temperature dependence of the
dielectric strength (Fig.~\ref{fig5}(c)). Starting from room temperature, the strength slightly increases but
significantly decreases below the onset of AFM and FE order. The decrease below $T_\mathrm{N}$ amounts to more
than 50\%. It is clear that this effect does not result from the change in the eigenfrequency, but must be
directly related to severe changes of the ionic plasma frequency which directly is related to the effective
charges involved in the specific phonon mode.\\

Further representative examples for the temperature dependencies of phonon excitations are shown in
Fig.~\ref{fig5}, documenting the temperature dependencies of the fitting parameters for  B$_{3u}$(2)
(Fig.~\ref{fig5}(d)-(f)) and B$_{3u}$(4) (Fig.~\ref{fig5}(g)-(i)). B$_{3u}$(2) exhibits a temperature dependence
of the eigenfrequency which is close to what is expected for a purely anharmonic behavior. However, the damping
reveals a slight increase below 100~K. The dielectric strength decreases below 100~K and starts to increase again
below the onset of AFM order. In contrast, mode B$_{3u}$(4) decreases in eigenfrequency below 100~K
(Fig.~\ref{fig5}(g)), a rather "normal" anharmonic temperature dependence of the damping constant
(Fig.~\ref{fig5}(h)) and on cooling from room temperature an unexpected and large increase of the polar strength
up to 100~K (Fig.~\ref{fig5}(i)). From this figure it becomes clear that there is no common behavior, but instead
eigenfrequencies, damping constants and polar strengths are different for each mode. A specific phonon mode
involves specific bonds of neighboring ions
which obviously are differently influenced by magnetic exchange as well as magnetoelectric interactions.\\

%\begin{figure}
%\includegraphics[width=0.9\columnwidth]{fig.5.eps}
%\caption{(Color online) Temperature dependencies of eigenfrequency, damping constant and mode strength for modes
%P2 (frames a-c) and P4 (frames d-f) in TbMnO$_3$ for \textbf{E}${}\parallel{}$\textbf{a} on semilogarithmic plots. The
%magnetic ordering temperatures are indicated by vertical dashed lines.} \label{fig5}
%\end{figure}

\begin{figure}
\includegraphics[width=\columnwidth]{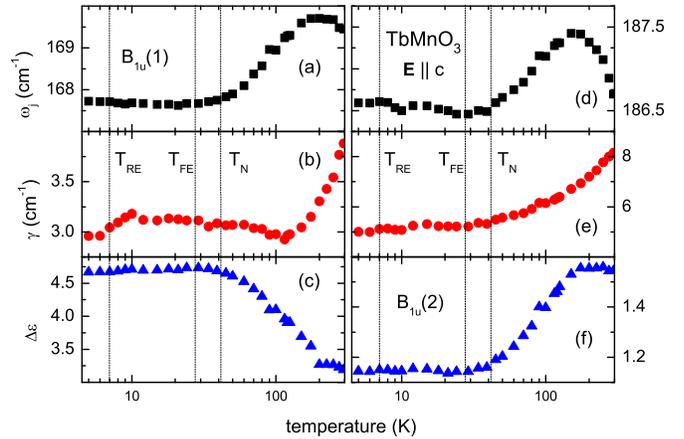}
\caption{(Color online) Temperature dependencies of eigenfrequency, damping constant and mode strength for modes
B$_{1u}$(1) and B$_{1u}$(2) in TbMnO$_3$ for \textbf{E}${}\parallel{}$\textbf{c} on semilogarithmic plots. The
magnetic ordering temperatures are indicated by vertical dashed lines.} \label{fig6}
\end{figure}

Similar observations can also be made for phonon excitations for \textbf{E}${}\parallel{}$\textbf{c}.
Representative examples are documented in Fig.~\ref{fig6}, where the left frames show the relevant fitting
parameters of mode B$_{1u}$(1) (Figs.~\ref{fig6}(a)-(c)) and the right frames those of mode B$_{1u}$(2)
(Figs.~\ref{fig6}(d)-(f)). Here, on decreasing temperature, both eigenfrequencies reveal a decrease below 100~K,
but the observed dipolar strengths of these two modes behave drastically different: on cooling we determined an
increase of mode B$_{1u}$(1) (Fig.~\ref{fig6}(c)), but a decrease for mode B$_{1u}$(2) (Fig.~\ref{fig6}(f)). The
effects amount in both cases to more than 10\% and seem not to be directly related to the AFM or FE phase
transitions. They start at
about 100~K, clearly above the appearance of any magnetic or polar order.\\

\begin{figure}
\includegraphics[width=0.9\columnwidth]{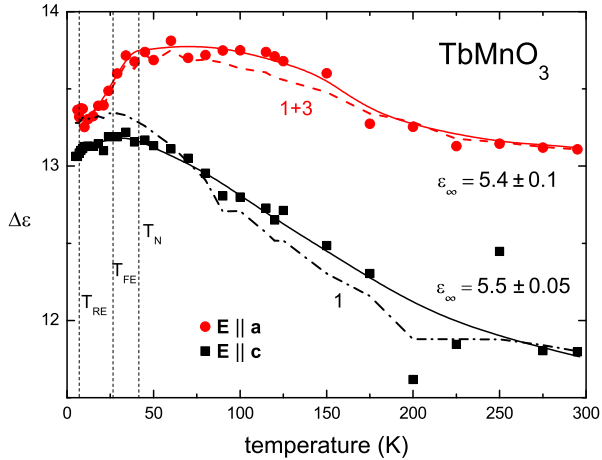}
\caption{(Color online) Temperature dependencies of the ionic contribution to the dielectric constant in TbMnO$_3$
for \textbf{E}${}\parallel{}$\textbf{a} (red dots) and \textbf{E}${}\parallel{}$\textbf{c} (black squares). The
magnetic ordering temperatures are indicated by vertical dashed lines. The electronic contributions to the
dielectric constant are also indicated. Solid lines are drawn to guide the eye. The sum of $\Delta\epsilon$ from
modes B$_{3u}$(1) and B$_{3u}$(2) for \textbf{E}${}\parallel{}$\textbf{a} is indicated as red dashed line.
$\Delta\epsilon$ of mode B$_{1u}$(1) for \textbf{E}${}\parallel{}$\textbf{c} is indicated as black dash-dotted
line. Both contributions have been shifted by a constant offset to agree with the values of the total dipolar
strengths at room temperature.} \label{fig7}
\end{figure}

Finally, we discuss the total strength of all polar active modes as documented in Table~\ref{tab1}, in order to
determine the overall temperature dependence of the dielectric constants along the crystallographic \textbf{a} and
\textbf{c} directions. Here, we summed over all modes. The temperature dependence of the total polar strength is
shown in Fig.~\ref{fig7}. For \textbf{E}${}\parallel{}$\textbf{a} the polar strength slightly increases and
considerably drops below the AFM and FE transition. This behavior provides experimental evidence that optical
weight is transferred from the phonon excitations to the electromagnons at lower wave numbers, which have not been
treated explicitly in the present work. It seems clear that the growing intensity of the electromagnons just
compensates for the loss of dipolar strength of the phonon modes. A closer inspection of the temperature
dependencies of all 9 B$_{3u}$ modes documents that this transfer of optical weight at the onset of non-collinear
spin order mainly results from mode B$_{3u}$(1) ($\Delta\epsilon = -0.25$; see Fig.~\ref{fig5}(c)) and mode
B$_{3u}$(3) ($\Delta\epsilon = -0.25$; not shown). It is interesting to note, that in clear contradiction to this
behavior, phonon mode B$_{3u}$(2) gains strength at low temperatures ($\Delta\epsilon = 0.2$; see
Fig.~\ref{fig5}(f)). It would be highly interesting to
find out, if in the latter case the transferred optical strength comes from spin waves via magnetoelectric interactions.\\

To document the fact that the temperature dependence of the dielectric constant in TbMnO$_3$ for
\textbf{E}${}\parallel{}$\textbf{a} mainly results from the modes B$_{3u}$(1) and B$_{3u}$(3), the sum of the
dipolar strength of these two modes is indicated in Fig.~\ref{fig7} by the red dashed line and has been shifted by
an offset to agree with the total dipolar strength of all modes at room temperature. This representation nicely
demonstrates that these two modes lose intensity below the onset of spiral spin order.

Most eigenfrequencies for \textbf{E}${}\parallel{}$\textbf{a} reveal rather normal anharmonic behavior, with a
smooth increase of the order of 1 or 2~cm$^{-1}$ on cooling. Significant spin-phonon coupling below the onset of
magnetic order only becomes apparent for mode B$_{3u}$(1), as documented in Fig.~\ref{fig3}(a). An unexpected
temperature dependence of the transverse optical modes has been deduced for modes B$_{3u}$(i) with i $=6,7,8$: In
all cases the eigenfrequencies pass through a maximum roughly located at 100~K and decrease on further cooling.
This decrease amounts to 2~cm$^{-1}$ for B$_{3u}$(6) and approximately 1~cm$^{-1}$ for modes B$_{3u}$(7) and
B$_{3u}$(8)
(not shown).\\

As shown in Fig.~\ref{fig7}, for \textbf{E}${}\parallel{}$\textbf{c} the total polar strength of all phonons and
concomitantly the dielectric constant shows a significant increase towards low temperatures and saturates below
100~K. No significant anomalies are observed on passing the magnetic phase transitions. The increase of polar
weight for $T>100$~K mainly results from mode B$_{1u}$(1) ($\Delta\epsilon = 1.5$; see Fig.~\ref{fig6}(c)). This
is indicated by the dash-dotted line which indicates the dielectric strength of this mode and has been shifted to
coincide with the sum over all dipolar strengths for this direction at room temperature.

The almost constant total weight at low temperatures (Fig.~\ref{fig7}) corresponds to the absence of transfer of
optical weight to spin excitations. Indeed, electromagnons in multiferroic rare earth manganites only appear with
polarization \textbf{E}${}\parallel{}$\textbf{a}.

For \textbf{E}${}\parallel{}$\textbf{c}, most eigenfrequencies exhibit rather canonical anharmonic behavior, with
the exception of modes B$_{1u}$(1) and B$_{1u}$(2) (see Figs.~\ref{fig6}(a) and (d)). For these modes the
eigenfrequencies pass through maxima at enhanced temperatures and decrease on further cooling. No anomalies show
up at the magnetic
transitions.\\

The total dielectric constant is the sum of all $\Delta\epsilon$ plus the electronic contribution. We determined
the electronic dielectric constant $\epsilon_\infty$ from fits to the reflectivity as function of temperature up
to 4,000~cm$^{-1}$ independently for both directions. We found no significant temperature dependence and arrived
at estimates for $\epsilon_\infty$ as 5.4 for \textbf{E}${}\parallel{}$\textbf{a} and 5.5 for
\textbf{E}${}\parallel{}$\textbf{c}.

This gives estimates of the dielectric constants of 18.7 for \textbf{E}${}\parallel{}$\textbf{a} and 18.6 for
\textbf{E}${}\parallel{}$\textbf{c} at low temperatures. One should have in mind that the dielectric constant for
\textbf{E}${}\parallel{}$\textbf{a} does not include the contribution of the electromagnon. For
\textbf{E}${}\parallel{}$\textbf{a} this result seems to be compatible with published values of measurements at
lower frequencies: Kimura \cite{Kimura2003} and Pimenov \cite{Pimenov2006} reported a dielectric constant
$\epsilon=24$ as measured at audio or GHz frequencies respectively. If we assume that most of the weight from the
120~cm$^{-1}$ phonon excitation is transferred to the 60~cm$^{-1}$ electromagnon (see e.g.
Ref.~\cite{Aguilar2009}), according to Fig.~\ref{fig6} we can estimate the contribution of the electromagnon to
the low-frequency dielectric constant to be of the order of 2 and there seems reasonable agreement between our
high-frequency and the low-frequency results. For \textbf{E}${}\parallel{}$\textbf{c}, Kimura reported values of
29 \cite{Kimura2003} or 23.5 \cite{Kimura2005}, both values are considerably enhanced compared to our result.

\section{Summary}

Using FIR spectroscopy we studied in detail the phonon excitations for \textbf{E}${}\parallel{}$\textbf{a} and
\textbf{E}${}\parallel{}$\textbf{c} and their coupling to the magnetic degrees of freedom of multiferroic
TbMnO$_3$. The main results can be summarized as
follows:\\

\renewcommand{\labelenumi}{\roman{enumi})}

\begin{enumerate}

\item \noindent
All nine B$_{3u}$ modes expected for \textbf{E}${}\parallel{}$\textbf{a} and the seven B$_{1u}$ modes expected for
\textbf{E}${}\parallel{}$\textbf{c} were assigned and listed in Table~\ref{tab1}. Almost all phonon modes reveal
anomalies in their eigenfrequencies, damping constants or ionic strengths. Most of these anomalies appear at the
magnetic and FE phase transitions indicating magnetoelectric coupling effects. However, in some cases anomalies
appear at significantly higher temperatures of about 100~K and may be linked to the Curie-Weiss temperature by
spin-phonon coupling similarly to other systems with magnetic frustration \mcite{Rudolf2007,
Rudolf2007b,*Hemberger2006,*Hemberger2007,*Rudolf2007a}. The effects in the eigenfrequencies are of the order of 1
to 2~cm$^{-1}$ in agreement with observations in other multiferroics \cite{Aguilar2006,Garcia-Flores2007}.
\item \noindent
The difference in frequency $\Delta \omega=\omega_{M_2}-\omega_{B_{3u}(1)}\cong$ 20~cm$^{-1}$ between the lowest
phonon B$_{3u}$(1) and the mode $M_2$, which emerges in the magnetically ordered phase, is almost constant with
temperature and corresponds to the energy scale of the low-energy electromagnons and antiferromagnetic resonances.
This may be an indication of a one-(electro)magnon + one phonon process.
\item \noindent
There is no apparent phonon softening close to the ferroelectric phase transition, an observation which is
compatible with the assumption that ferroelectric order is driven by the onset of helical spin order and the
ferroelectricity is of improper type.
\item \noindent
There is transfer of spectral weight of phonons with \textbf{E}${}\parallel{}$\textbf{a} to electromagnon
excitations, i.e.
 spin waves which gain polar weight through strong magnetoelectric coupling
(see Fig.~\ref{fig3}~b). In total this transfer of dielectric strength corresponds to a decrease of the dielectric
constant just below the phonon frequencies of approximately 0.5 (see Fig.~\ref{fig7}).
\item \noindent Unexpectedly, we observed a significant increase of the
dielectric constant for \textbf{E}${}\parallel{}$\textbf{c} which results from an increase of the effective ionic
plasma frequency on cooling. This observation could be a characteristic scenario for spin-driven ferroelectrics.
While in displacive ferroelectrics polar order is established by a continuous softening of a transverse optical
phonon mode with a concomitant increase of the static dielectric constant, it seems that in the spin-driven
improper ferroelectrics an increase of the dielectric constant follows from an increase of dynamical charges
indicating charge transfer processes. Recent density functional theory \cite{Xiang2008, Malashevich2008} and first
principal calculations \cite{Picozzi2007} stress the importance of conventional polar atomic displacements, but
also demonstrate the importance of purely electronic contributions to the polarization, which results from a
spin-orbit interaction modifying the hybridization of electronic orbitals \cite{Katsura2005}.
\end{enumerate}

\begin{acknowledgement}
We acknowledge fruitful discussions with A.~Pimenov and Th.~Kopp. This research was supported by the Deutsche
Forschungsgemeinschaft (DFG) via the Collaborative Research Center, SFB484 (Augsburg)
\end{acknowledgement}

\bibliography{Publikation_tbmno3_vs7}

\begin{thebibliography}{43}
\bibitem{Note}
Note that the excitation M$_1$ was not included in the fit routine, because it
  is very close to the experimental low-frequency limit and a
  band-fourier-filter was used to smoothen the interference fringes
  superimposed on the spectra due to multiple reflections in the sample.
\end{thebibliography}
\bibliographystyle{epj}

\end{document}